\documentstyle[prl,aps,epsfig,multicol]{revtex}
\begin{document}
\draft

%%%%%%%%%%%%%%%%%%%%%%%%%%%%%%%%%%%%%%%%%%%%%%%%%%%%%%%%%%%%%%%
% Title
%

\title{Suppressed $\pi^0$ Production at Large Transverse Momentum 
in Central Au+Au Collisions at $\sqrt{s_{_{NN}}}$~=~200~GeV }

% NOTE:  Brant will add the author list later.  Its length
% does not count toward the length estimate.  Instead, 24 
% lines are counted toward the length to approximate a 
% typical set of several authors and a few institutions.
%
% Use the following 5 lines exactly as written:

%% Author and Institution List for RUN-2

\author{
S.S.~Adler,$^{5}$
S.~Afanasiev,$^{17}$
C.~Aidala,$^{5}$
N.N.~Ajitanand,$^{43}$
Y.~Akiba,$^{20,28}$
J.~Alexander,$^{43}$
R.~Amirikas,$^{12}$
L.~Aphecetche,$^{45}$
S.H.~Aronson,$^{5}$
R.~Averbeck,$^{44}$
% ...
\\(PHENIX Collaboration)\\
}
\address{
% ...
$^{5}$Brookhaven National Laboratory, Upton, NY 11973-5000, USA\\
% ...
$^{12}$Florida State University, Tallahassee, FL 32306, USA\\
% ...
$^{17}$Joint Institute for Nuclear Research, 141980 Dubna, Moscow Region, Russia\\
% ...
$^{20}$KEK, High Energy Accelerator Research Organization, Tsukuba-shi, Ibaraki-ken 305-0801, Japan\\
% ...
$^{28}$LPC, Universit{\'e} Blaise Pascal, CNRS-IN2P3, Clermont-Fd, 63177 Aubiere Cedex, France\\
% ...
$^{43}$Chemistry Department, Stony Brook University, SUNY, Stony Brook, NY 11794-3400, USA\\
$^{44}$Department of Physics and Astronomy, Stony Brook University, SUNY, Stony Brook, NY 11794, USA\\
$^{45}$SUBATECH (Ecole des Mines de Nantes, CNRS-IN2P3, Universit{\'e} de Nantes) BP 20722 - 44307, Nantes, France\\
% ...
}

\date{\today}        
\maketitle

%%%%%%%%%%%%%%%%%%%%%%%%%%%%%%%%%%%%%%%%%%%%%%%%%%%%%%%%%%%%%%%
% Abstract
%

\begin{abstract}
Transverse momentum spectra of neutral pions in the range  1 $< p_T <$ 10~GeV$/c$ 
have been measured at mid-rapidity by the PHENIX experiment at RHIC in 
Au+Au collisions at $\sqrt{s_{_{NN}}}$ = 200~GeV. The $\pi^0$ 
multiplicity in central reactions is significantly below the yields measured at the 
same $\sqrt{s_{_{NN}}}$ in peripheral Au+Au and $p+p$ reactions scaled by the  
number of nucleon-nucleon collisions. For the most central bin, the suppression 
factor is $\sim$2.5 at $p_T$ = 2 GeV$/c$ and increases to $\sim$4-5 at $p_T\approx$ 4 GeV$/c$. 
At larger $p_T$, the suppression remains constant within errors. The deficit is 
already apparent in semi-peripheral reactions and increases smoothly with centrality.
\end{abstract}

\begin{multicols}{2}   % This is needed only for the "multicols" style
\narrowtext            % This is needed only for the "multicols" style
%%%%%%%%%%%%%%%%%%%%%%%%%%%%%%%%%%%%%%%%%%%%%%%%%%%%%%%%%%%%%
%NOTES: 
%
%1.  For PRL do not use section headings.
%
%2.  Do not worry about indenting the first line of a paragraph.  Just
%    insert a blank line between paragraphs.  Similarly, if you want 
%    an equation to stay within a paragraph, do not put a blank line
%    before or after the equation.
%
%3.  Do not imbed figures or tables; place them all at the end (see below).
%
%4.  Name all references and use "\cite{refname}" in the text to cite them.
%    (The RevTeX macro will replace this with "[1]" in proper PR style.)
%
%5.  The list of references must be ordered in the same sequence as they
%    occur in the text.
%
%6.  Use our standard aknowldegement below as the last paragraph of your
%    text.  (Yes, it does count toward the length!).
%
%%%%%%%%%%%%%%%%%%%%%%%%%%%%%%%%%%%%%%%%%%%%%%%%%%%%%%%%%%%%%
% general introduction
%
% \marginpar{{\small \em Intro}}
%
%\twocolumn

High energy collisions of heavy ions provide the means to study
Quantum Chromodynamics (QCD) at energy densities where lattice 
calculations~\cite{latt02} predict a transition from hadronic matter to a 
deconfined, chirally symmetric plasma of quarks and gluons (QGP). 
The large center-of-mass energies, $\sqrt{s_{_{NN}}}\approx$ 200~GeV, available 
in Au+Au collisions at the Relativistic Heavy Ion Collider (RHIC), 
have resulted in a significant production of high transverse momentum hadrons 
($p_T>$ 2 GeV$/c$) for the first time in heavy-ion physics. High $p_T$ particle 
production in hadronic collisions results from the fragmentation of quarks and 
gluons emerging from the initial high $Q^2$ parton-parton scatterings~\cite{owens}. 
Thus, hard processes in nucleus-nucleus ($AA$) collisions provide direct information 
on the early partonic phases of the reaction.
In the absence of nuclear medium effects, hard scattering yields 
in $AA$ reactions are expected to scale like an incoherent superposition 
of nucleon-nucleon ($NN$) collisions because of the small 
probability of hard scattering processes per $NN$ collision. 
In $AA$ reactions, the number of $NN$ collisions ($N_{coll}$) at impact parameter $b$  
is simply proportional to the geometric nuclear overlap function, $T_{AA}(b)$, and
can be calculated in an eikonal approach~\cite{glauber}. %In this way, 
After scaling by the nuclear geometry, spectra of high $p_T$ particles measured 
in $AA$ reactions can be compared to the baseline ``vacuum'' ($p+p$, $p+\bar{p}$) 
and ``cold medium'' ($p,\, d+A$) data, as well as to 
perturbative~\cite{vitev,ina,levai,xnwang} and ``classical''~\cite{dima} 
QCD predictions. Any departure from the expected $N_{coll}$-scaled result
provides information on the strongly interacting medium in central heavy-ion 
reactions.

One of the most significant observations from the first RHIC run (Run-1) was the 
suppressed yield of moderately high $p_T$ neutral pions ($p_T\approx$ 1.5 - 4.0 GeV$/c$) 
in central Au+Au at $\sqrt{s_{_{NN}}}$ = 130 GeV with respect to the 
$N_{coll}$-scaled $p+p$ and peripheral Au+Au data~\cite{ppg003}. 
This result points to strong medium effects present in central Au+Au 
and has triggered extensive theoretical studies on its 
origin~\cite{vitev,ina,levai,xnwang,dima,mueller,arleo,carlos-urs,gallmeister}.
Most of these studies are based on the prediction \cite{Gyu90,BDMPS} that a 
QGP would induce multiple gluon radiations from the scattered fast partons, 
effectively leading to a suppression of high $p_T$ hadronic fragmentation 
products (``jet quenching''). Alternative interpretations have been proposed 
based on initial-state gluon saturation~\cite{dima} 
or final-state hadronic interactions~\cite{gallmeister}.

This Letter presents $\pi^0$ results obtained by the PHENIX experiment
in Au+Au collisions at $\sqrt{s_{_{NN}}}$ = 200~GeV and compares them to 
the $p+p\rightarrow\pi^0\,X$ data measured in the same experiment at the
same center-of-mass energy~\cite{ppg024}.
During the 2001-2002 RHIC run (Run-2), PHENIX obtained $\pi^0$ data measured in the 
electromagnetic calorimeter (EMCal). The present analysis uses 30 million minimum bias events, 
triggered by a coincidence between the Zero Degree Calorimeters (ZDC) and the Beam-Beam 
Counters (BBC), with vertex position $|z|<$ 30 cm. 
In Run-2, the PHENIX EMCal was fully instrumented providing a 
total solid angle coverage at mid-rapidity of approximately $\Delta\eta = 0.7$ and 
$\Delta \phi = \pi$ and the total collected $\pi^0$ statistics was a factor of $\sim 100$ 
larger than in Run-1~\cite{ppg003}. The combination of larger acceptance, high statistics, 
and the measurement of $p+p$ data in the same detector permits a precise 
study of the high $p_T$ $\pi^0$ production mechanisms in $AA$ collisions at RHIC.

Neutral pions are reconstructed via their $\pi^0\rightarrow\gamma\gamma$ decay 
through an invariant mass analysis of $\gamma$ pairs detected in the EMCal~\cite{nim_emc} 
which consists of 6 lead-scintillator (PbSc) and 2 lead-glass \v{C}erenkov (PbGl) sectors. 
The large radial distance of the calorimeters to the interaction region ($>$ 5~m)
and their fine granularity ($\Delta\eta\times\Delta\phi\approx$ 0.01$\times$0.01) 
keep the tower occupancy low, $<$10\% even in the highest multiplicity Au+Au 
events. The energy calibration is obtained from beam tests and, in the case of PbSc, 
from cosmic rays data and the known minimum ionizing energy peak of charged hadrons 
traversing the calorimeter. It is then confirmed using the $\pi^0$ mass, as well as the 
agreement of the calorimeter energy with the measured momentum of identified 
electrons. The systematic error on the absolute energy scale is less than 1.5\%. 
Photon-like clusters are identified in the EMCal by applying %particle identification (PID) 
time-of-flight and shower profile cuts~\cite{nim_emc}. 
The selected clusters are binned in pair invariant mass $m_{\gamma \gamma}$ and 
$p_T$. An additional energy asymmetry cut, 
$|E_{\gamma 1}-E_{\gamma 2}|/(E_{\gamma 1}+E_{\gamma 2})<$
0.7(PbGl), 0.8(PbSc), is applied to the reconstructed pairs.
The signal-to-background in peripheral (central) is approximately 20 (5) and 0.5 (0.01)
for the highest and lowest $p_T$, respectively.
The combinatorial background is estimated and subtracted by mixing clusters from different events 
with similar centrality and vertex, and normalizing the distribution in a region outside 
the $\pi^0$ mass peak. The $\pi^{0}$ yield in each $p_{T}$ bin is determined by 
integrating the subtracted $m_{\gamma \gamma}$ distribution in a $\pm3\sigma$ window 
determined by a $p_T$-dependent parameterization of Gaussian fits to the $\pi^{0}$ peaks.

The raw PbSc and PbGl $\pi^0$ spectra are normalized to one unit of rapidity and
full azimuth (this acceptance correction rises quickly with $p_T$ up to
a $\sim$1/0.35 pure geometric factor). The spectra are further corrected for 
(i) the detector response (energy resolution, dead areas), 
(ii) the reconstruction efficiency (analysis cuts), and (iii) the occupancy effects 
(cluster overlaps). These corrections are quantified by embedding simulated single 
$\pi^0$'s from a full PHENIX GEANT~\cite{geant} simulation into real events,
and analyzing the merged events with the same analysis cuts used 
to obtain the real yields. Each correction is determined, for each centrality bin, 
as the ratio of the input (simulated) to the reconstructed $p_T$ distribution.
The overall yield correction amounts to $\sim$2.5 with a centrality dependence
$\lesssim$25\%. The losses are dominated by fiducial and asymmetry cuts.

The main sources of systematic errors in the PbSc and PbGl measurements 
are due to the uncertainties in: (i) the yield extraction, (ii) the yield correction, 
and (iii) the energy scale. The relative contributions of these effects to 
the total error differ for the PbSc and PbGl (Table~\ref{tab:syst}).
The weighted average of the two independent measurements reduces the total error. 
The nominal energy resolution \cite{nim_emc} is adjusted in the simulation to reproduce 
the true width of the $\pi^0$ peak observed at each $p_T$, smearing the energies with a 
constant term of 7\% for PbSc and $\sim$9\% for PbGl. The shape, position, and width
of the $\pi^0$ peak measured in all different centralities are then confirmed to be well 
reproduced by the embedded data. %The uncertainties in the photon-like cluster selection are 
%evaluated comparing the results of different PID cuts. 
The final systematic errors on the spectra are at the level of $\sim$10\% at 1 GeV$/c$ and $\sim$17\% at the highest 
$p_T$ (Table~\ref{tab:syst}). A correction for the true mean value of the $p_T$ bin 
is applied to the steeply falling spectra. No corrections have been applied to account
for the contribution of feed-down $\pi^0$'s (mainly coming from $K^0_s$ and $\eta$ decays) 
which are $<$5\% based on HIJING \cite{hijing} simulations. 

%%%%%%%%%%%%%%%%%%%%%%%%%%%%%%%%%%%%%%%%%%%%%%%%%%%%%%%%%%%%%%%%%%%%%%%%
% Results and discussion
%

\begin{figure}[ht]
%\begin{center}
\centerline{\epsfig{file=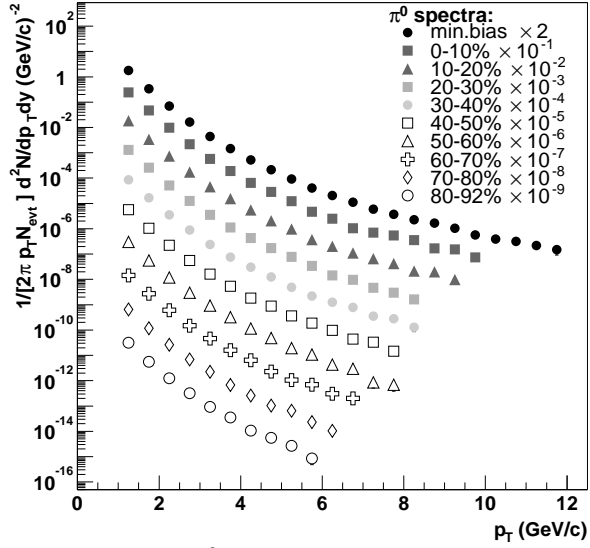,width=0.9\linewidth}}
\caption{Invariant $\pi^0$ yields at mid-rapidity as a function of $p_T$ for 
minimum bias and 9 centralities in Au+Au at $\sqrt{s_{_{NN}}}$ = 200~GeV 
[0-10\% (80-92\%) is most central (peripheral)].}
%(Table~\protect\ref{tab:Ncoll}).}
%\end{center}
\label{fig:pt_spectra} 
\end{figure}

The event centrality is determined by correlating the charge detected
in the BBC with the energy measured in the ZDC detectors. A Glauber model 
Monte Carlo (MC) combined with a simulation of the BBC and ZDC responses~\cite{ppg001,foot0,foot1} 
gives an estimate of the associated number of binary collisions ($N_{coll}$) 
and participating nucleons ($N_{part}$) in each centrality bin (Table~\ref{tab:Ncoll}).
The fully corrected and combined PbSc and PbGl $\pi^0$ $p_T$ distributions 
are shown in Fig.~\ref{fig:pt_spectra} for minimum bias and for 9 centrality 
bins scaled by factors of 10 for clarity.

We quantify the medium effects on high $p_T$ production in $AA$ collisions with the 
{\it nuclear modification factor} given by the ratio of the measured $AA$ invariant yields 
to the $NN$ collision scaled $p+p$ invariant yields:
\begin{equation} 
R_{AA}(p_T)\,=\,\frac{(1/N^{evt}_{AA})\,d^2N^{\pi^0}_{AA}/dp_T dy}{\langle N_{coll}\rangle/\sigma_{pp}^{inel} \,\times\, 
d^2\sigma^{\pi^0}_{pp}/dp_T dy},
\label{eq:R_AA}
\end{equation}
where the $\langle N_{coll}\rangle/\sigma_{pp}^{inel}$ is just 
the average Glauber nuclear overlap function, $\langle T_{AuAu} \rangle$, 
in the centrality bin under consideration (Table~\ref{tab:Ncoll}). 
$R_{AA}(p_T)$ measures the deviation of $AA$ data from an incoherent superposition 
of $NN$ collisions. %in terms of suppression ($R_{AA}<$1) or enhancement ($R_{AA}>$1). 
For $p_T\lesssim$ 2 GeV$/c$, $R_{AA}$ is known to be below unity, since the bulk of 
particle production is due to soft processes which scale closer to
the number of participant nucleons \cite{ppg001} than to $N_{coll}$.

\begin{figure}[ht]
\centerline{\epsfig{file=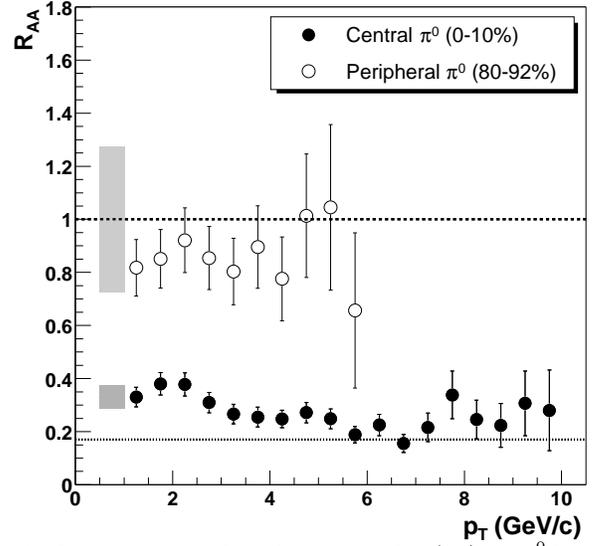,width=0.9\linewidth}}
\caption{Nuclear modification factor $R_{AA}(p_T)$ for $\pi^0$ in central 
(closed circles) and peripheral (open circles) Au+Au at $\sqrt{s_{_{NN}}}$ 
= 200~GeV.  The error bars include all point-to-point experimental 
($p+p$, Au+Au) errors. The shaded bands represent the 
fractional uncertainties in $\langle T_{AuAu} \rangle$ and in the $\pi^0$ yields 
normalization added in quadrature, which can move all the points up or down together
(in the central case the shaded band shown is the fractional error for the first point).}
\label{fig:R_AA} 
\end{figure}

Figure~\ref{fig:R_AA} shows $R_{AA}$ as a function of $p_T$ 
for $\pi^0$ measured in 0-10\% central (closed circles) and 80-92\% peripheral 
(open circles) Au+Au. The PHENIX $p+p\,\rightarrow\,\pi^0$ data~\cite{ppg024} is used as the 
reference in the denominator. The $R_{AA}$ values for central collisions are noticeably below unity, 
as found at 130 GeV~\cite{ppg003}, and in contrast to the enhanced high $p_T$ $\pi^0$
production ($R_{AA}>$1) observed at CERN-SPS energies~\cite{wa98} and
interpreted in terms of initial-state $p_T$ broadening effects (``Cronin effect''~\cite{cronin}).
Within errors, peripheral Au+Au collisions behave like a superposition of $p+p$ collisions 
with regard to high $p_T$ $\pi^0$ production ($R_{AA}\approx$ 1). In central collisions, the suppression is 
smallest at 2 GeV$/c$ and increases to an approximately constant suppression factor of 
1/$R_{AA}\approx$~4-5 over the $p_T$ range of 4 - 10 GeV$/c$, $\sim$30\% above the expectation 
from $N_{part}$ scaling (dotted line in Fig.~\ref{fig:R_AA}).

The magnitude and $p_T$ dependence of $R_{AA}$ (corresponding to parton fractional 
momenta $x\approx 2p_T/\sqrt{s}\sim 0.02-0.1$ at midrapidity), is inconsistent 
with the expectations of leading-twist ``shadowing'' effects on the nuclear parton 
distribution functions alone~\cite{eks98-vogt}. 
Different jet quenching calculations~\cite{vitev,ina,levai,xnwang,mueller,arleo,carlos-urs}, 
based on medium-induced radiative energy loss, can reproduce the magnitude %{\it magnitude} 
of the $\pi^0$ suppression assuming the formation of a hot and dense partonic system. 
The predicted $p_T$ dependence of the quenching, however, varies in the different models. 
All models that include the Landau-Pomeranchuck-Migdal (LPM) interference effect~\cite{BDMPS,glv} predict 
$R_{AA}$ effectively $\propto \sqrt{p_T}$~\cite{mueller}. Such a trend is not compatible 
with our data over the entire $p_T$ range. %Other approaches, such 
%as constant energy loss per parton scattering, are also not supported
%as discussed in~\cite{ina}. 
Analyses which combine LPM jet quenching together 
with shadowing and initial-state $p_T$ broadening generally reproduce the whole 
$p_T$ dependence of the $\pi^0$ suppression~\cite{vitev}, as do recent approaches 
that take into account detailed balance between parton emission and absorption~\cite{xnwang}. 
However, based solely on the data presented here, we are not able to distinguish between 
partonic or hadronic~\cite{gallmeister} energy loss scenarios.

%%%%%%%%%%%%%%%%%%%%%%%%%%%%%%%%%%%%%%%%%%%%%%

\begin{figure}[ht]
%\vspace{-0.9cm}
\centerline{\epsfig{file=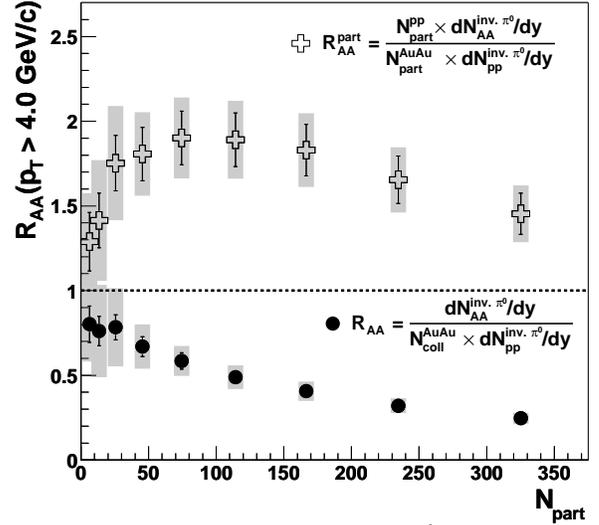,width=0.9\linewidth}}
%\vspace{0.cm}
\caption{Ratio of Au+Au over $p+p$ $\pi^0$ yields integrated above 4~GeV$/c$ and 
normalized using $N_{coll}$ (circles) and $N_{part}$ (crosses), 
as a function of centrality given by $N_{part}$. The errors bands and bars are the same as for 
Fig.~\protect\ref{fig:R_AA}.}
\label{fig:R_AA_vs_cent}
\end{figure}

The centrality dependence of the high $p_T$ $\pi^0$ suppression is shown in
Fig.~\ref{fig:R_AA_vs_cent} as a function of $N_{part}$. The suppression is characterized as 
the ratio of Au+Au over $p+p$ yields integrated above 4~GeV/{\it c} and normalized 
using two different scalings. $R_{AA}$ (circles) denotes the $N_{coll}$ scaling as in 
Eq.~(\ref{eq:R_AA}), whereas $R_{AA}^{part}$ (crosses) indicates $N_{part}$ scaling 
expected in scenarios dominated either by gluon saturation~\cite{dima} or by surface emission 
of the quenched jets~\cite{mueller}. Fig.~\ref{fig:R_AA_vs_cent} indicates %(lower side)
that the transition from the $N_{coll}$ scaling behaviour
($R_{AA}\sim$~1) apparent in the most peripheral region, to the strong suppression 
seen in central reactions ($R_{AA}\sim$~0.25) is smooth. In addition, although there is
no exact participant scaling ($R_{AA}^{part}>$ 1 for all centralities),
the $\pi^0$ production per participant pair above 4~GeV/{\it c} is approximately 
constant over a wide range of intermediate centralities, in qualitative agreement 
with a parton saturation model prediction~\cite{dima}.

%%%%%%%%%%%%%%%%%%%%%%%%%%%%%%%%%%%%%%%%%%%%%%%%%%%%%%%%%%%%%%%%%
% Conclusion
%
%\marginpar{{\small \em Concl}}

In summary, transverse momentum spectra of neutral pions have been measured at 
mid-rapidity up to $p_T\approx$ 10~GeV$/c$ for 9 centrality bins
of Au+Au collisions at $\sqrt{s_{_{NN}}}$ = 200~GeV. The spectral shape and 
invariant yield for peripheral reactions are consistent with those of $p+p$ 
reactions scaled by the average number of inelastic $NN$ collisions. 
Central yields, on the other hand, are significantly lower than peripheral 
Au+Au and $p+p$ scaled yields, as found at $\sqrt{s_{_{NN}}}$ = 130 GeV. 
The observed suppression increases slowly with $p_T$ to as much as a factor 
4-5 in the 10\% most central collisions, remaining constant within errors 
above $\sim$4~GeV/$c$. The suppression is already apparent in semi-peripheral 
reactions and increases smoothly with centrality. The magnitude of the deficit 
can be reproduced by parton energy loss calculations in an opaque medium, 
but its $p_T$ and centrality dependence puts strong constraints 
on the details of energy loss and the properties of the medium.
The role of initial-state effects, including shadowing, 
$p_T$ broadening, and gluon saturation will be studied with data
from the recent RHIC run using d+Au, where final-state medium effects such 
as jet quenching are absent.

%%%%%%%%%%%%%%%%%%%%%%%%%%%%%%%%%%%%%%%%%%%%%%%%%%%%%%%%%%%%
% Acknowledgements
%

%\section{Acknowledgements}

%\section{Acknowledgements}   % Run-2 short form for PRL

We thank the staff of the Collider-Accelerator and Physics
Departments at BNL for their vital contributions.  We acknowledge
support from the Department of Energy and NSF (U.S.A.), MEXT and
JSPS (Japan), CNPq and FAPESP (Brazil), NSFC (China), CNRS-IN2P3
and CEA (France), BMBF, DAAD, and AvH (Germany), OTKA (Hungary), 
DAE and DST (India), ISF (Israel), KRF and CHEP (Korea),
RAS, RMAE, and RMS (Russia), VR and KAW (Sweden), U.S. CRDF 
for the FSU, US-Hungarian NSF-OTKA-MTA, and US-Israel BSF.

%REFERENCES:  Use \begin{references} and \end{references}.  Do not use
%             \begin{thebibliography} and \end{thebibliography}.
%             You may either 
%		(a) enter all citations explicitly or 
%               (b) use some "\def" shorthand notations.
%             Our first paper used approach (a) and our second used (b).
%             Here are the two reference lists as examples of how to proceed:

\begin{table}
\caption{
Summary of the dominant sources of systematic errors on the PbSc and PbGl
$\pi^0$ yields, and total errors on the combined measurements.  
The error ranges are quoted for the lowest to highest $p_{T}$ values.
For the combined $\pi^0$ spectra and $R_{AA}$, we quote the 
approximate i) statistical, ii) point-to-point systematic, 
and iii) absolute normalization (for most central and peripheral reactions) errors.}
\begin{tabular}[hb]{l|cccc}
Source   	 & & syst. error PbSc & \multicolumn{2}{c}{syst. error PbGl} \\\hline
Yield extraction & & 10\%	    	     & \multicolumn{2}{c}{6 - 7\%} \\ 
Yield correction & &  8\%	    	     & \multicolumn{2}{c}{8\%} \\
Energy scale     & &  3 - 11\% 	     & \multicolumn{2}{c}{7 - 13\%} \\\hline\hline
Total error (\%) &	stat.     & syst.	&  \multicolumn{2}{c}{normalization} \\
		 &		     & 			&  central  & peripheral \\\hline
Spectra     	 &	2 - 40       & 10 - 17		&  5        &   5 \\
$R_{AA}$    	 &	2 - 45       & 11 - 22		& 14        &  30\\
\end{tabular}
\label{tab:syst}
\end{table}

\begin{table} 
\caption{
Centrality bin, average nuclear overlap function, number of $NN$
collisions, and number of participant nucleons obtained from a Glauber
MC \protect\cite{foot0,foot1} and the BBC and ZDC responses 
for Au+Au at $\sqrt{s_{NN}}$ = 200~GeV. The centrality bin is expressed as percentiles of
$\sigma_{AuAu}$ = 6.9 b. The last line refers to minimum bias
collisions.}
\begin{tabular}[]{cccc}
Centrality & $\langle T_{AuAu}\rangle$ (mb$^{-1}$) & $\langle N_{coll}\rangle$ & $\langle N_{part}\rangle$\\ \hline
 0-10\%   &  22.75 $\pm$ 1.56  &  955.4 $\pm$ 93.6  &  325.2 $\pm$  3.3 \\
10-20\%   &  14.35 $\pm$ 1.00  &  602.6 $\pm$ 59.3  &  234.6 $\pm$ 4.7 \\
20-30\%   &   8.90 $\pm$ 0.72  &  373.8 $\pm$ 39.6  &  166.6 $\pm$ 5.4 \\
30-40\%   &   5.23 $\pm$ 0.44  &  219.8 $\pm$ 22.6  &  114.2 $\pm$ 4.4 \\
40-50\%   &   2.86 $\pm$ 0.28  &  120.3 $\pm$ 13.7  &   74.4 $\pm$ 3.8 \\
50-60\%   &   1.45 $\pm$ 0.23  &   61.0 $\pm$  9.9  &   45.5 $\pm$ 3.3 \\
60-70\%   &   0.68 $\pm$ 0.18  &   28.5 $\pm$  7.6  &   25.7 $\pm$ 3.8 \\
60-80\%   &   0.49 $\pm$ 0.14  &   20.4 $\pm$  5.9  &   19.5 $\pm$ 3.3 \\
60-92\%   &   0.35 $\pm$ 0.10  &   14.5 $\pm$  4.0  &   14.5 $\pm$ 2.5 \\
70-80\%   &   0.30 $\pm$ 0.10  &   12.4 $\pm$  4.2  &   13.4 $\pm$ 3.0 \\
70-92\%   &   0.20 $\pm$ 0.06  &    8.3 $\pm$  2.4  &    9.5 $\pm$ 1.9 \\
80-92\%   &   0.12 $\pm$ 0.03  &    4.9 $\pm$  1.2  &    6.3 $\pm$ 1.2 \\
min. bias &   6.14 $\pm$ 0.45  &  257.8 $\pm$ 25.4  &  109.1 $\pm$ 4.1 \\
\end{tabular}
\label{tab:Ncoll}
\end{table}

%FIGURES:  Place all the figures here (after the references) in sequence.

%%%%%%%%%%%%%%%%%%%%%%%%%%%%%%%%%%%%%%%% Figure 1.

%\clearpage
%%%%%%%%%%%%%%%%%%%%%%%%%%%%%%%%%%%%%%%% Figure 2.
\vspace{5cm}
%\vspace{-0.5cm}

%%%%%%%%%%%%%%%%%%%%%%%%%%%%%%%%%%%%%%%% Figure 3.

\end{multicols}


\begin{references}
\bibitem{foot0}The Run-2 minimum bias trigger is inefficient for very peripheral reactions 
and records $92.2^{+2.5}_{-3.0}$\% of $\sigma_{AuAu}$.
\end{references}
\end{document}